\documentclass[apl,twocolumn,superscriptaddress,preprintnumbers,floatfix]{revtex4}
\usepackage[pdftex]{graphicx}
\usepackage{amssymb}
\usepackage{array}
\usepackage{verbatim}
\usepackage[ansinew]{inputenc}
\usepackage[paperwidth=216mm,paperheight=280mm,centering,hmargin=2.54cm,vmargin=2.54cm]{geometry}
\usepackage[english]{babel}
\usepackage{setspace}
\usepackage{subfig}
\usepackage{natbib}
\begin{document}
\DeclareGraphicsExtensions{.pdf,.png,.jpg,.eps,.tiff}
\title{Increased surface flashover voltage in microfabricated devices}
\author{R. C. Sterling}
\affiliation{Department of Physics and Astronomy, University of Sussex, Brighton, BN1 9QH, UK}
\author{M. D. Hughes}
\affiliation{Department of Physics and Astronomy, University of Sussex, Brighton, BN1 9QH, UK}
\author{C. J. Mellor}
\affiliation{School of Physics and Astronomy, University of Nottingham, University Park, Nottingham, NG7 2RD, UK}
\author{W. K. Hensinger\footnote{w.k.hensinger@sussex.ac.uk}}
\affiliation{Department of Physics and Astronomy, University of Sussex, Brighton, BN1 9QH, UK}
\email{W.K.Hensinger@sussex.ac.uk}
\begin{abstract}
With the demand for improved performance in microfabricated devices, the necessity to apply greater electric fields and voltages becomes evident. When operating in vacuum, the voltage is typically limited by surface flashover forming along the surface of a dielectric. By modifying the fabrication process we have discovered it is possible to more than double the flashover voltage. Our finding has significant impact on the realization of next-generation micro- and nano-fabricated devices and for the fabrication of on-chip ion trap arrays for the realization of scalable ion quantum technology.
\end{abstract}
\maketitle
\thispagestyle{headings}
Microfabricated devices such as microelectromechanical systems (MEMS) operating in vacuum have a multitude of applications. These include space applications such as nanoelectrospray thruster arrays for spacecraft \cite{Paine2004112,krpoun,4443818,4801975} and spacecraft solar arrays \cite{Velez,6084761,4663173}, to earth bound applications such as field emitter arrays. Most recently they have become a crucial tool for the realization of quantum technologies based on ion traps. Ion traps have proven themselves to be a powerful tool for many experiments in modern science. They exhibit good isolation from the surrounding environment and long coherence times are achievable \cite{Dwineland}. As a result, ion trap experiments have been used to explore cavity QED \cite{KellerNature,CQED}, the measurement of frequency standards \cite{LoriniClock}, quantum simulators \cite{QsimRev, IslamQsim} and quantum information processing \cite{PhysRevLett.74.4091, Dwineland, Hffner2008155}. The Paul ion trap has been used to demonstrate unparalleled success towards the implementation of the first scalable quantum computer, meeting most of the requirements for qubit control, and extensive work is being carried out towards a scalable architecture within which to store and control the qubits \cite{hughes}.

However, there still remain many challenging technical issues to address before a fully scalable ion trap quantum computer can be built. Not least is building an architecture within which thousands of ions may be stored, shuttled and manipulated. Recent work has focused on using microfabrication techniques to build ion trap arrays, harnessing the massive parallelism and accuracy achievable with modern semiconductor fabrication facilities \cite{hughes}. This has lead to many advancements such as state manipulation from integrated microwave waveguides \cite{muwavegates} and integrated optical fibers \cite{kim:214103}. Despite these exciting advances, there still remain several fundamental problems with microfabricated traps. In order to allow for sufficient trap depths and large secular frequencies in microfabricated ion traps that feature large ion-electrode distances, the ability to apply large voltages is required. Such voltages would require large separations between electrodes and result in exposed dielectrics. This leaves the ion susceptible to any uncontrolled charges collected on these exposed dielectrics. These effects will have a slow time dependance and make effective long term compensation troublesome. In order to minimize exposed regions of dielectric, electrodes are fabricated with only small gaps, on the order of several micrometers. Alternatively dielectrics can be shielded completely from the ion using multi-layered geometries, but again, microfabrication considerations limit layer thicknesses to a few micrometers. This results in large electric fields between electrodes and if proper care is not taken electrical breakdown can occur, destroying the chip. Electrical breakdown in vacuum via a connecting surface is known as surface flashover.

Additionally, the close proximity of the ion to the electrode surface induces anomalous heating of the ion's motional state, which scales approximately as $d^{-4}$, where $d$ is the ion-electrode separation \cite{PhysRevA.61.063418}. There have been several techniques demonstrated recently which manage to suppress heating by performing surface cleaning \cite{in-situ-cleaning, laser_cleaning} or operating at cryogenic temperatures \cite{PhysRevLett.100.013001}, but additional improvements can be made by designing traps with an increased ion-electrode separation. This also has benefits in easing optical alignment across the trap surface, reducing unwanted laser scatter from trap electrodes and reducing the effect of uncontrolled charging of dielectrics and electrodes \cite{wang:104901}.

For these reasons, high-fidelity operations are more difficult with microfabricated ion traps as they currently lack the benefits afforded to macroscopic traps. It is therefore desirable to find ways in which microfabricated ion trap arrays can be optimized in order not only to improve their functionality, as seen in Refs. \cite{hughes, muwavegates, kim:214103, AminiScalable}, but also allow for larger voltages to be applied. This would allow for larger ion-electrode distances and smaller electrode-electrode spacings. In this letter, we present a simple method to significantly increase the voltage that can be applied to MEMS and microfabricated devices in general and on-chip ion trap arrays in particular. By increasing the maximum voltage before surface flashover occurs, traps can be designed with increased ion-electrode separations and smaller spacings between adjacent electrodes.

No studies have been published on how to improve surface flashover voltage in microfabricated devices. In fact, only one experiment has been carried out characterizing surface flashover voltages at relevant electrode separations of between 5 $\mu$m and 20 $\mu$m. This experiment was carried out for a particular dielectric material with a particular fabrication and cleaning process \cite{878383}. Therefore, we first investigate the difference between static and rf breakdown and then show how the choice of dielectric allows to significantly increase the breakdown threshold.

Electrical breakdown in vacuum, also known as surface flashover, is described by secondary electron emission avalanche (SEEA) across the dielectric surface. Electrons hop across the dielectric surface, which desorbs gas molecules from the surface leading to a Townsend-like breakdown through this gas layer \cite{neuber:3084,pillai:2983,anderson:1414}. This is a function of the amount of desorbed gas per unit area at the point of flashover, $M_{cr}$, the electron emission and impact energies, $A_0$ and $A_1$ respectively, the efficiency of electron stimulated gas desorption, $\gamma$, molecule ejection velocity, $v_0$, electron velocity, $v_e$, which is given by $v_e=5.94\times10^5 \sqrt{A_1}$ m/s \cite{pillai:2983}, and $\theta$, which is the angle that electrons are emitted from the triple point. The point at which the dielectric, cathode and vacuum meet, given by $\tan\theta=[2A_0/(A_1-A_0)]^{\frac{1}{2}}$ \cite{pillai:2983}. The flashover voltage is given by \cite{pillai:2983}
\begin{equation}
V_b=\Bigg[\frac{\varphi de}{2\epsilon_0}\Bigg]^{\frac{1}{2}}
\label{flashoverV2}
\end{equation}
where
\begin{displaymath}
\varphi =\frac{v_0M_{cr}A_1}{\gamma v_e\tan\theta},
\end{displaymath}
where $d$ is the electrode separation, $e$ is the electron charge and $\epsilon_0$ is the permittivity of free space.

Unfortunately, there is only a limited amount of information regarding $M_{cr}$, $\gamma$ and $v_0$, with measurements ranging over several orders of magnitude for different experimental setups \cite{Avdienko1977643,neuber:3084,pillai:2983,anderson:1414}, and $A_1$ is only known for a handful of common dielectrics \cite{pillai:146}. Therefore we will treat $\varphi$ as a fitting parameter to compare between static and rf measurements and between different fabrication processes.

In order to set a base line for surface flashover, test samples were fabricated using a common, simple fabrication technique of gold electrodes deposited onto quartz. The electrodes were deposited by e-beam evaporation, depositing a chromium seed layer followed by a 500 nm layer of gold. Electrodes were patterned using standard photolithography and formed using wet etching. The electrodes were separated by gaps from 3 to 15 $\mu$m, in 2 $\mu$m steps. The test chips were superglued to a ceramic chip carrier, and connections made by wire bonding 30 $\mu$m gold wire between the electrodes and chip carrier. The chip carrier was attached to a high power vacuum feedthrough and mounted inside a glass belljar, then the system was evacuated using a turbomolecular pump to a pressure of $\approx5\times10^{-4}$ Pa. Negative static voltage was applied by attaching a 5 kV supply to the feedthrough with an in built voltage divider, supplying a 0-10 V monitor voltage that could be measured by a calibrated voltmeter with an error of $\pm$ 10 mV, resulting in an experimental error of $\pm$ 5 V. RF voltage was applied by attaching a 2 $\mu$H inductor to the feedthrough, forming a resonant LCR circuit with the chip. The resonant frequency of the inductor-chip circuit of 22.0 $\pm$ 0.5 MHz. A 30 W amplifier was connected to the inductor via a bidirectional coupler with a capacitive probe measuring the voltage applied to the sample. Breakdown was measured by slowly ramping up the voltage while observing the sample through a lens. Upon flashover a bright plasma discharge develops and the voltage ramp is stopped. The voltage is then recorded, the error of this measurement is $\pm$ 7 \% for both static and rf measurements.

\begin{figure}
\centering
\includegraphics[scale=0.22]{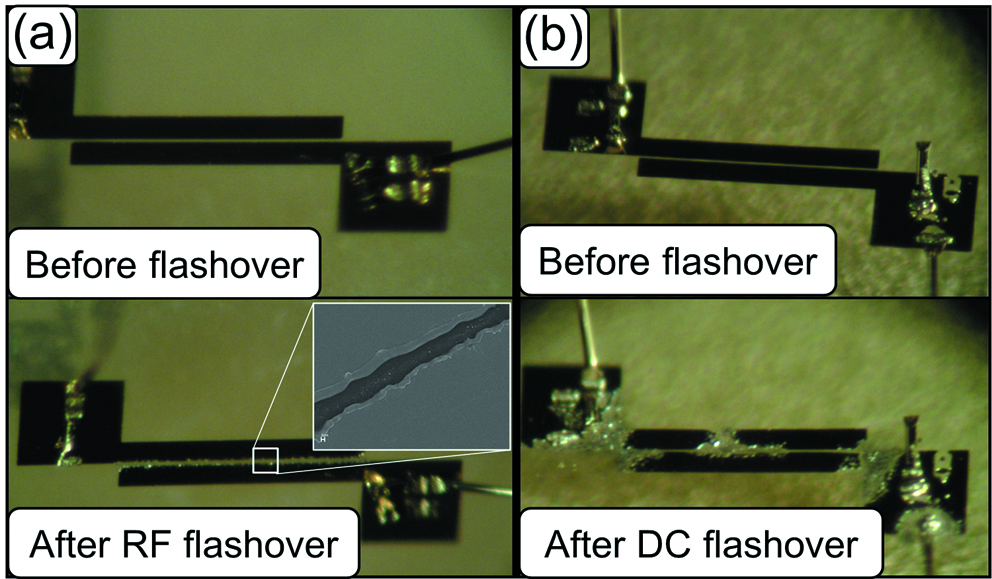}
\caption{Samples before and after a flashover measurement is taken. (a) Picture showing a sample with 7 $\mu$m electrode spacing before and after RF flashover occurred. The insert shows an electron microscope image of the damaged electrodes. (b) Picture showing a sample with 7 $\mu$m electrode spacing before and after static flashover occurred.}
\label{beforeandafter}
\end{figure}

Figure \ref{beforeandafter} shows four microscope images of test samples. Fig. \ref{beforeandafter}(a) shows a 7 $\mu$m gap before and after rf surface flashover, the inset shows an electron microscope image of the damaged electrodes. Figure \ref{beforeandafter}(b) shows a 7 $\mu$m sample before and after static flashover occurred. A significant visual difference can be observed. For rf flashover, the closest edge along the full length of the electrodes has been eroded until flashover can no longer be sustained. This differs from static breakdown which occurs at the sharp edges of the electrodes where the E-field is strongest. Upon breakdown there is a sudden reduction in impedance and a rapid discharge of capacitively stored charge, leading to large portions of the electrodes being destroyed during flashover. There are a number of mechanisms that may prevent such damage for rf flashover. Plasma dissipation during the low voltage periods in the oscillation, when the electric field switches polarity may limit this damage. Another explanation relates to the Q-factor of the resonant rf circuit. When flashover occurs, the resulting resistive component of the LC resonator circuit will rapidly lower the Q of the RF resonator and may stop the discharge. However once the discharge is stopped, the Q increases again and so flashover re-occurs.

\begin{figure}
\centering
\includegraphics[scale=0.17]{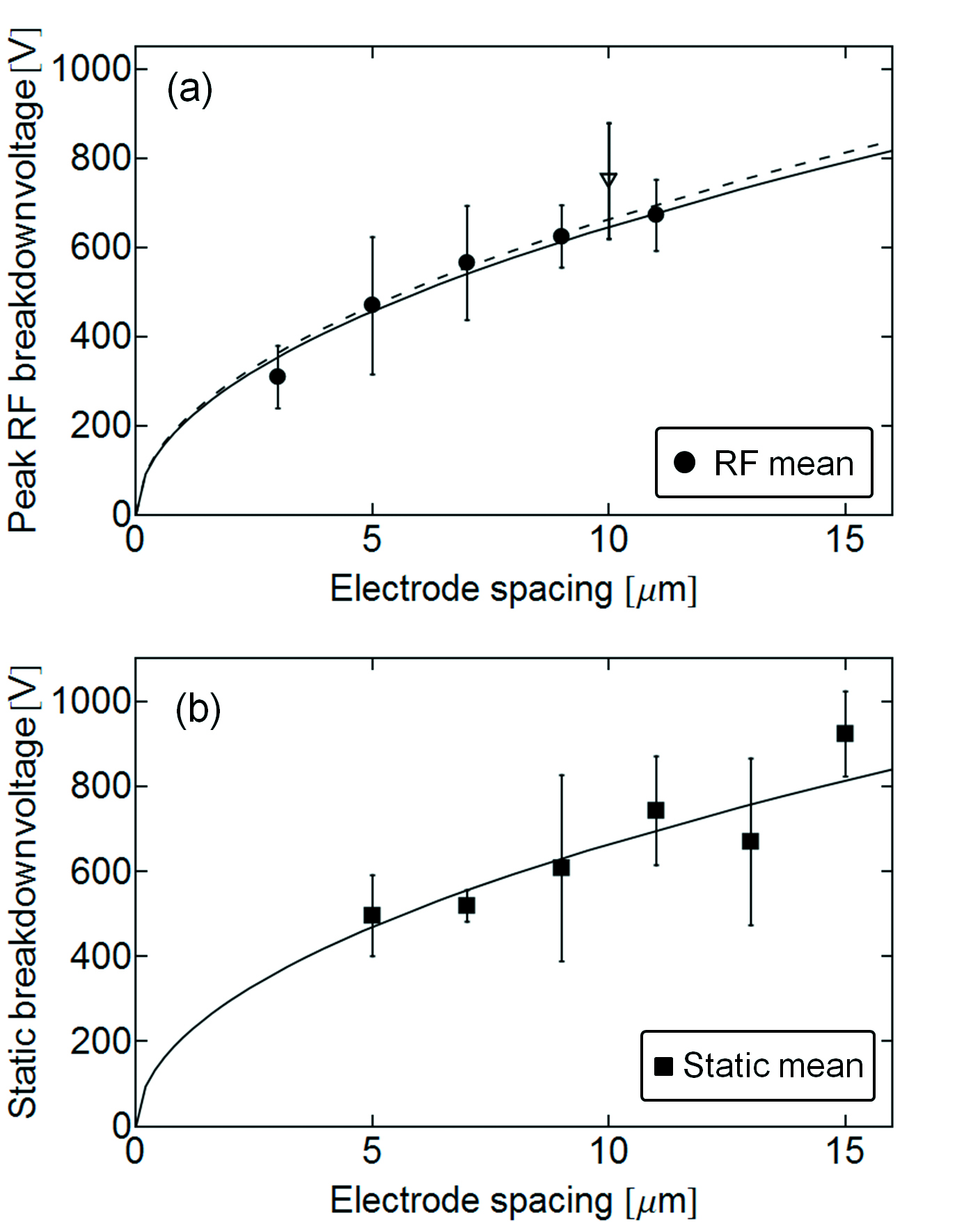}
\caption{(a) The mean flashover voltage is shown as solid circles, the error bars correspond to the standard deviation. Equation \ref{flashoverV2} is fitted to the data with $\varphi_{rf}=4.6\times10^{18}$ eV m$^{-2}$ also shown is the fitted line for static flashover as a dashed line. The average flashover voltage from electrodes fabricated using a lift-off process is shown as an empty inverted triangle. (b) Static flashover voltage data is shown as empty squares, the mean flashover voltage is shown as solid squares. Equation \ref{flashoverV2} is fitted to the data with $\varphi_{dc}=4.9\times10^{18}$ eV m$^{-2}$.}
\label{BreakdownRF}
\end{figure}

The results for both static and rf flashover are shown in Fig. \ref{BreakdownRF}, along with a plot of Equ. \ref{flashoverV2} using $\varphi$ as a fitting parameter, the voltage magnitude is plotted for both rf and static measurements. The error bars correspond to the standard deviation, excluding the static measurement at 15 $\mu$m where only one point was measured, in this case the measurement error is given.

Equation \ref{flashoverV2} was fitted for both the rf and static flashover data, giving $\varphi_{rf}=4.6\times10^{18}$ eV m$^{-2}$ and $\varphi_{dc}=4.9\times10^{18}$ eV m$^{-2}$. This represents a difference of $\approx$ 5 \% between rf and static flashover voltage, showing no statistically significant difference in breakdown voltage, despite the visual differences.

We also compare our measurements to silicon dioxide deposited using low pressure chemical vapor deposition on a silicon wafer \cite{878383}. These measurements show an improvement over our quartz measurements, with $\varphi=16\times10^{18}$ eV m$^{-2}$. We speculate that this discrepancy compared to our measurements is a result of an oxygen plasma etch prior to the measurement, as such an etch will remove remaining organic materials from the surface. It has been reported that during flashover, outgassed materials from the samples are predominantly CO, CO$_2$ and H$_2$ \cite{42158}. This is confirmed with our own residual gas analyzer measurements, in which we also observe large peaks at these molecules. Removing the majority of these compounds will lower $\gamma$ in Equ. \ref{flashoverV2} and therefore increase the flashover voltage.

In multilayer fabrication processes, flashover will occur across deposited dielectrics and not the substrate itself. The surface properties of the deposited dielectric are likely to differ from that of a polished quartz wafer, so it is important to compare the flashover voltage for both bulk and deposited dielectrics. Samples were prepared on a quartz substrate with a deposited, layered dielectric structure. The layered structure was chosen to prevent the formation of pinholes in the deposited dielectric, these pinholes may reduce bulk breakdown between surface electrodes and buried conductors, which are often present in microfabricated ion traps and other microfabricated devices. The layered dielectric consisted of three layers of alternating 100 nm of aSiO and 72 nm aSiN, with an additional layer of either aSiO or aSiN on top. The dielectrics were deposited using plasma enhanced chemical vapor deposition (PECVD) at 250$^\circ$C, in an isothermal PECVD reactor (Corial D250). Rather than using e-beam evaporation and photolithography (as was for the data shown in Fig. \ref{BreakdownRF}), the electrodes were formed using a lift-off process and thermally evaporating an adhesion layer of titanium and then 200 nm of gold. To ensure there was no change in flashover as a result of the different electrode fabrication process, additional measurements across a quartz wafer were performed with electrodes formed in this way. This is shown in Fig. \ref{BreakdownRF}(a) as an inverted empty triangle, illustrating that the surface flashover voltage does not depend on the gold deposition process.

\begin{figure}
\centering
\includegraphics[scale=0.17]{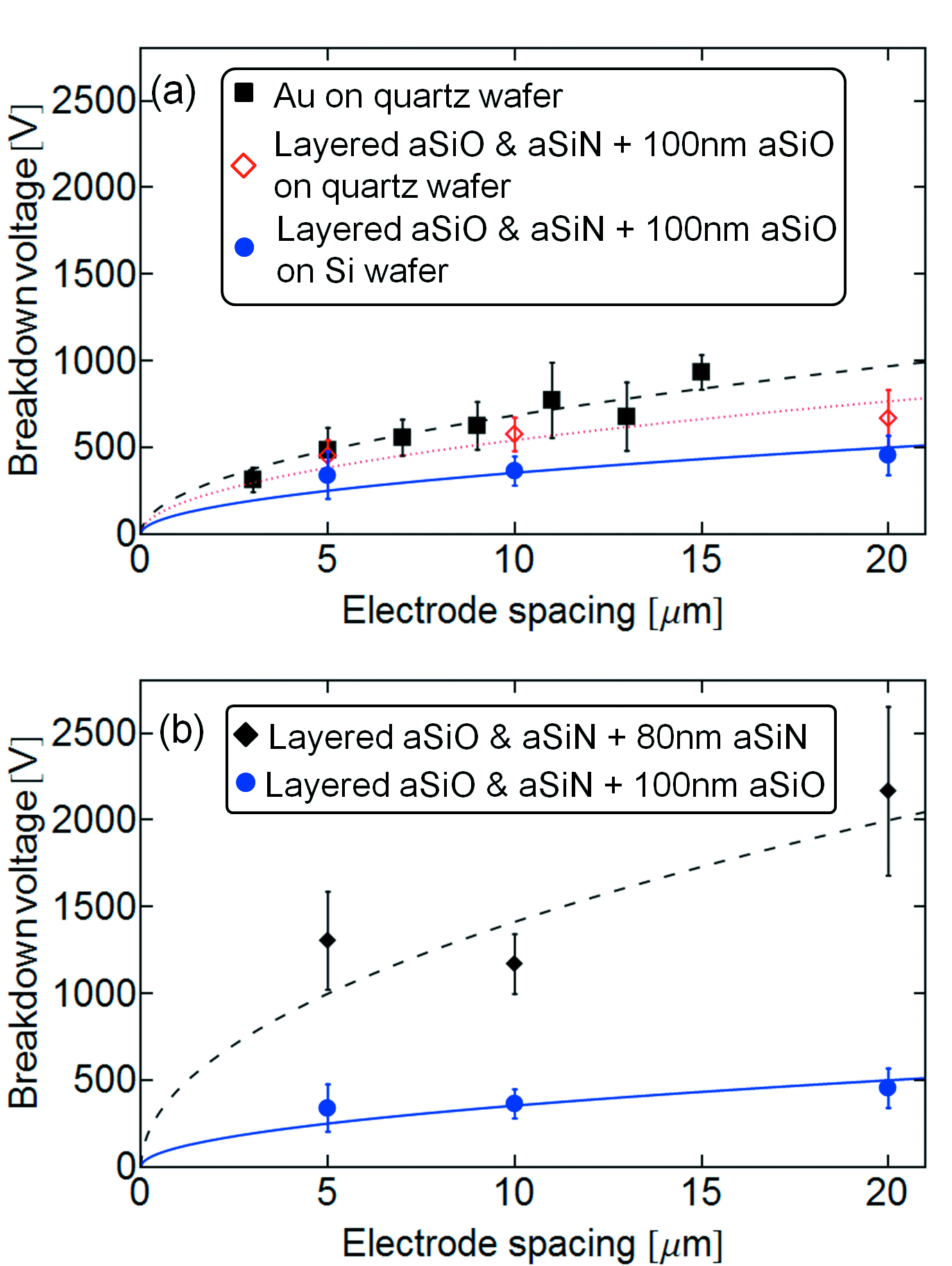}
\caption{Two graphs comparing flashover voltage between test samples fabricated using different fabrication processes and materials. (a) The mean flashover voltage across quartz is shown by solid black squares. This is the mean of both the rf and static data. Also shown is the flashover voltage across a aSiO surface layer on top of a multi-layered aSiO and aSiN deposition, shown by empty red diamonds. There is a modest reduction in flashover voltage but within the error bars of our measurements. Flashover measurements across aSiO on a silicon substrate are shown as blue circles, showing $\approx$ 30 \% reduction in flashover voltage. (b) A comparison between flashover across a surface layer of aSiO, shown by blue circles, and aSiN, shown by solid black diamonds, both on a layered dielectric on a silicon substrate. There is a $\approx$ 3.6 fold improvement in flashover voltage when using aSiN as a surface layer.}
\label{comparison1}
\end{figure}

Flashover measurements across the deposited dielectric were performed at 5, 10 and 20 $\mu$m. The mean of these measurements is shown in Fig. \ref{comparison1}(a) as empty red diamonds. We also show the mean of both the rf and static quartz flashover measurements, as black squares. The flashover across a deposited layered dielectric with an aSiO surface shows a slight reduction but are within the error bars of the gold on quartz measurements.

Using dielectrics deposited by PECVD allows for the use of a conductor beneath, such as a metallic ground plane, buried electrodes or a conductive substrate. The flashover voltage primarily depends on the surface properties of the dielectric: the electron impact energy, gas desorption efficiency and the amount of adsorbed gas. Therefore the bulk property of the dielectric is not taken into account in Equ. \ref{flashoverV2}, as it is assumed to be uniform. However, when there is a conducting or semi-conducting material just below the surface, no electric field lines will pass through this conductor. This implies that there will be a higher density of electric field lines near the electrode. We performed numerical simulations of the electrodes using boundary element method (BEM) software (Charged Particle Optics, by Electronoptics), and found that the electric field lines are indeed highly concentrated near the high-voltage electrode. The result of introducing a ground plane is that the electric field deviates significantly from the uniform electric field typically assumed in SEEA models \cite{neuber:3084,pillai:2983,anderson:1414}. This modification of the electric field increases the electric field magnitude near the triple point and introduces an electric field perpendicular to the dielectric surface. To investigate how a conducting substrate or ground plane effects flashover, measurements were performed on PECVD aSiO deposited on a silicon wafer. These measurements are shown in Fig. \ref{comparison1}(a) as blue circles, the flashover voltage for samples deposited on a silicon substrate show a reduction of $\approx$ 30 \% with $\varphi=1.4\times10^{18}$ eV m$^{-2}$ when compared to aSiO on a quartz substrate. This is expected considering the higher electric field amplitude we have found in our BEM electric field simulations with a conductive substrate.

In order to increase the voltage when flashover occurs, we have investigated other dielectric materials and fabrication processes, discovering it is possible to substantially increase the voltage at which flashover occurs. We have found that silicon nitride offers a significant improvement in flashover performance compared to silicon dioxide. Silicon nitride is a common alternative to silicon dioxide, and is readily available in microfabrication cleanrooms, making it a convenient substitution. Samples were prepared on a silicon substrate in a similar manner to the layered aSiO samples, however with a surface layer of 80 nm of aSiN. A comparison between flashover on aSiO and aSiN is shown in Fig. \ref{comparison1}(b), with aSiO shown as blue circles and aSiN as black diamonds. This shows that there is a significant improvement in flashover voltage when using aSiN instead of aSiO as a dielectric. We find $\varphi=18.1\times10^{18}$ eV m$^{-2}$, corresponding to an increase in the flashover voltage across aSiN by a factor of approximately 3.6 compared to aSiO. When carrying out measurements across 0.5 $\mu$m of aSiN on a quartz substrate we have observed values of $\varphi$ as large as $29.9\times10^{18}$ eV m$^{-2}$, however with a large spread of the data points resulting in a larger uncertainty in predicted surface flashover voltage.

We note that further improvements in surface flashover voltages may be possible with aSiN incorporating an oxygen plasma etch.

We have demonstrated that by using silicon nitride instead of silicon dioxide as a dielectric in microfabricated devices, the flashover voltage in vacuum can be significantly improved. Additionally we have shown that surface flashover is slightly affected by the substrate material.

Our analysis as to how surface flashover voltage can be improved has not only step-changing applications for ion trap arrays but also for other microfabricated and MEMS devices operating in vacuum, such as nanoelectrospray thruster arrays for spacecraft \cite{Paine2004112,krpoun,4443818,4801975} and spacecraft solar arrays \cite{Velez,6084761,4663173}, where high electric fields are desired. Further improvements may also be found by adjusting sample preparation prior to testing. $\varphi$ is a function of not only dielectric material but also the density of adsorbed gas molecules. Processes to lower this density may further increase flashover voltage, including oxygen plasma etches \cite{878383} or in vacuum cleaning as demonstrated recently using an argon ion beam \cite{in-situ-cleaning} or laser ablation \cite{laser_cleaning}. These results offer opportunities for the improvement of surface ion trap technology towards scalable architectures. Our results show that the separation between electrodes can be reduced, in principle, by an order of magnitude by incorporating the findings of our work in the fabrication process. This also greatly reduces potentially exposed dielectric surface area in ion trap arrays. Our results are also very promising for many other microfabrication applications such as MEMS, NEMS, quantum devices, field emitter arrays and space technology.

\begin{acknowledgments}
The authors would like to thank David Brown and Keith Schwab for fabricating the first gold on quartz test samples and James Siverns, Simon Webster and Sebastian Weidt for helpful discussions. We would also like to acknowledge the support of the UK Engineering and Physical Sciences Research Council (EP/E011136/1, EP/G007276/1), European Community's Seventh Framework Programme (FP7/2007-2013) under grant agreement no. 270843 (iQIT), the European Commission’s Sixth Framework Marie Curie International Reintegration Programme (MIRG-CT-2007-046432), the Nuffield Foundation and the University of Sussex.
\end{acknowledgments}


\end{document}